\documentclass[a4paper,12pt,]{article}
\pdfoutput=1 %

\usepackage[T1]{fontenc}
\usepackage[latin9]{inputenc}
\usepackage[a4paper]{geometry}
\usepackage{verbatim}
\usepackage{amsmath}
\usepackage{stmaryrd}
\usepackage{hyperref}

\usepackage{amsmath}
\usepackage{amssymb}

\usepackage{amsmath} 
\usepackage{url} 
\usepackage{epsfig} 
\usepackage{cite} 
\usepackage{psfrag} 

\usepackage{xcolor}

\newcommand{\nn}{\nonumber}

 \newcounter{RSQ}

\definecolor{ForestGreen}{HTML}{6C6E41}
\definecolor{ForestBlue}{HTML}{3E6169}
\definecolor{ForestOrange}{HTML}{E58035}
\definecolor{ForestBrown}{HTML}{813726}

\usepackage{slashed}

\begin{document} 
\begin{flushright}
MITP/23-050\\

\end{flushright}

\vskip3cm
\begin{center}
{{\Large \bf Endpoint divergences in inclusive $\bar B \to X_s \gamma$}\footnote{\footnotesize Based on talks given by R.S. at the 20th annual workshop on Soft-Collinear Effective Theory (SCET 2023), Berkeley, March 27-30, 2023 
and by T.H. at  21st Conference on Flavor Physics and CP Violation (FPCP 2023), Lyon, May 29 - June 2, 2023 and on Ref. [17].}}

\end{center}

 \vspace{1.5cm}
\begin{center}
{\sc Tobias Hurth$^{a}$ and Robert~Szafron,$^{b}$} 
\\[6mm]
{\it ${}^a$PRISMA+ Cluster of Excellence and Institute of Physics (THEP),\\
Johannes Gutenberg University, D-55099 Mainz, Germany}\\[6mm]
{\it ${}^b$Department of Physics, Brookhaven National Laboratory, Upton, N.Y., 11973, U.S.A.}
\\[2.3cm]

\end{center}

\begin{abstract}
The subleading so-called resolved contributions represent the largest uncertainty in the inclusive decay 
mode $\bar B \to X_s \gamma$. However there had been no complete proof of factorization of these subleading  contributions. This  failure of factorisation can be  traced back to  endpoint divergences and cured by recently proposed refactorisation techniques.
\end{abstract}

\newpage


\newpage

\section{Introduction}

The inclusive decay modes $\bar B \to X_{s,d} \gamma$ and  $\bar  B \to X_{s,d}  \ell^+\ell^-$ are golden modes of the indirect search for new physics at the Belle-II experiment~\cite{Hurth:2003vb,Hurth:2010tk,Hurth:2012vp}. The Belle II experiment will collect two orders of magnitude larger data samples  than the $B$ factories~\cite{Belle-II:2018jsg} and will measure the inclusive decays with the highest precision. Therefore, subleading effects have become  now relevant on the theory side. 

Within the heavy mass expansion (HME), these inclusive so-called penguin modes are dominated by the partonic contributions, which can be calculated perturbatively, and sub-leading contributions start at the quadratic level,                
$(\Lambda/ m_b)^2$ only.  However, it is well known that this operator product expansion breaks down in these inclusive modes if one considers operators beyond the leading ones. This breakdown manifests in nonlocal power corrections, also called resolved contributions.  They are characterised by  containing  subprocesses in which the photon couples to light partons instead of connecting directly to the effective weak-interaction vertex~\cite{Lee:2006wn}. 

These resolved contributions can be systematically calculated using soft-collinear effective theory (SCET). In the case of the inclusive $\bar B \to X_{s} \gamma$ decay, all resolved contributions to $O(1/m_b)$ have been calculated  some time ago~\cite{Benzke:2010js}. Also, the analogous contributions to the inclusive 
$\bar B \to X_{s ,d} \ell^+ \ell^-$ decays have been analysed to $O(1/m_b)$~\cite{Hurth:2017xzf,Benzke:2017woq}. 

Recently, the uncertainty due to the resolved contribution was reduced with the help of a new hadronic input~\cite{Gunawardana:2019gep,Benzke:2020htm}. But these resolved contributions still represent the largest uncertainty of order $4-5\%$ in the prediction of the inclusive decay rate $\bar B \to X_s \gamma$~\cite{Misiak:2015xwa} and of the low-$q^2$ observables of 
$\bar B \to X_{s ,d} \ell^+ \ell^-$~\cite{Huber:2015sra,Huber:2019iqf,Huber:2020vup}.

Moreover, a large scale dependence and also a large charm mass dependence were identified in the lowest order result of the resolved contribution, which calls for a systematic calculation of $\alpha_s$ corrections and renormalisation group (RG) summation~\cite{Benzke:2020htm,Benzke:2022cbw}. A mandatory prerequisite for this task is an all-order in the strong coupling constant $\alpha_s$ factorisation formula for the  subleading power corrections.

One finds a  factorisation formula for the various  contributions to the inclusive penguin 
decays~\cite{Benzke:2010js}, where the symbol $\otimes$ denotes the convolution of the soft and jet functions.
\begin{align}\label{fact2}
   &d\Gamma(\bar B\to X_s\,  \gamma, \ell^+ \ell^-)
   = \sum_{n=0}^\infty\,\frac{1}{m_b^n}\, \sum_i\,H_i^{(n)} J_i^{(n)}\otimes S_i^{(n)} \nonumber \\
   &\qquad + \sum_{n=1}^\infty\,\frac{1}{m_b^n}\,\bigg[ \sum_i\,H_i^{(n)} J_i^{(n)}\otimes S_i^{(n)}\otimes\bar J_i^{(n)}
    + \sum_i\,H_i^{(n)} J_i^{(n)}\otimes S_i^{(n)} \otimes\bar J_i^{(n)}\otimes\bar J_i^{(n)} \bigg] \,. 
\end{align}
The first line describes the so-called direct contributions, while the second line contains the resolved contributions. The latter appear first only at the order $1/m_b$ in the heavy-quark expansion.  
Here hard functions $H_i^{(n)}$ describe physics at the high scale $m_b$.
$J_i^{(n)}$ are the so-called jet functions which represent the physics of the hadronic final state $X_s$ at the intermediate hard-collinear scale $\sqrt{m_b \Lambda_{\rm QCD}}$. The soft functions $S_i^{(n)}$, the so-called shape functions, parameterise the hadronic physics at  the scale $\Lambda_\text{QCD}$. Within the resolved contributions, we have a new ingredient in the factorisation formula, the so-called anti-hardcollinear  jet functions 
$\bar J_i^{(n)}$ due to the coupling of virtual photons with virtualities of order $\sqrt{m_b \Lambda_\text{QCD}}$ to light partons instead of the weak vertex directly. They are not represented by cut propagators as the usual jet functions but as full propagator functions dressed by Wilson lines. 

However, the specific resolved ${O}_{8g} - {O}_{8g}$ contribution does not factorise because the convolution integral is ill-defined. The authors of Ref.~\cite{Benzke:2010js} claimed that there is an essential difference between divergent convolution integrals in power-suppressed contributions of exclusive $B$ decays and the divergent convolution integral in the present case, while the former were of IR origin, the latter divergence were of UV nature. Nevertheless, using a hard cut-off in the resolved contribution,  the sum of direct and resolved ${O}_{8g} - {O}_{8g}$ contributions  was shown to be scale and scheme independent at the lowest order. However the failure of factorisation did not allow for a consistent resummation of large logarithms.  In a recent  paper, the divergences in the resolved and in the direct contributions were identified as endpoint divergences. It was shown that also the divergence in the direct contribution can be traced back to a divergent convolution integral~\cite{Hurth:2023paz}. 

Recently new techniques~\cite{Liu:2019oav,Liu:2020wbn,Beneke:2020ibj,Beneke:2022obx} were presented in specific collider applications, which allow  for an operator-level reshuffling of terms within the factorisation formula so that all endpoint divergences cancel out. This  idea of refactorisation  was now implemented in this flavour example of the resolved contributions, which includes nonperturbative soft functions, and the subleading shape functions,  not present in collider applications~\cite{Hurth:2023paz}. This is the first QCD application of these new refactorsiation techniques in flavour physics. 

In the following, we present the various steps of this analysis. This analysis leads to   a renormalised factorisation theorem on the operator level for these resolved contributions  to all orders in the strong coupling constant. This result establishes the validity of the general factorisation theorem, given in Eq.~\ref{fact2}, - also for the  ${O}_{8g} - {O}_{8g}$ contributions. This theorem now allows for higher-order calculations of the resolved contributions and consistent summation of large logarithms~\cite{Hurth:2023paz}.

\section{Matching on SCET and degeneracies in the EFT} 

\begin{figure}
\begin{center}
\includegraphics[width=0.4\textwidth]{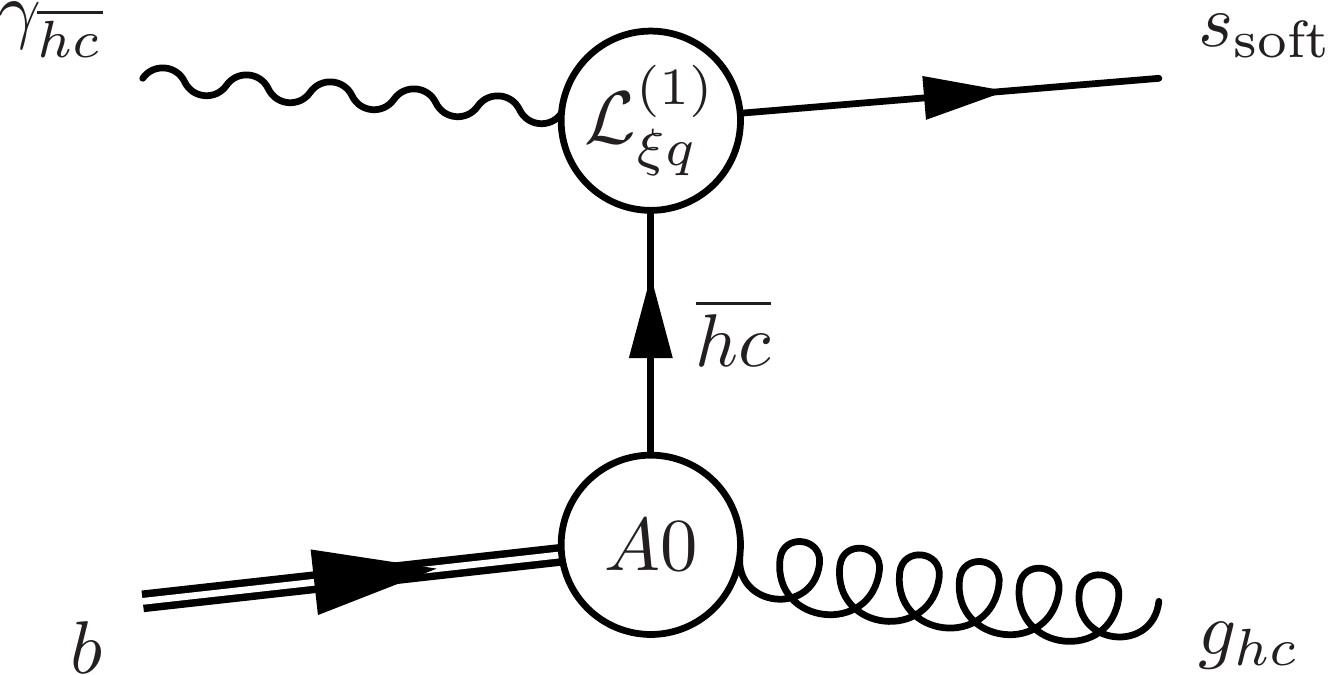} \hspace{1cm} \includegraphics[width=0.4\textwidth]{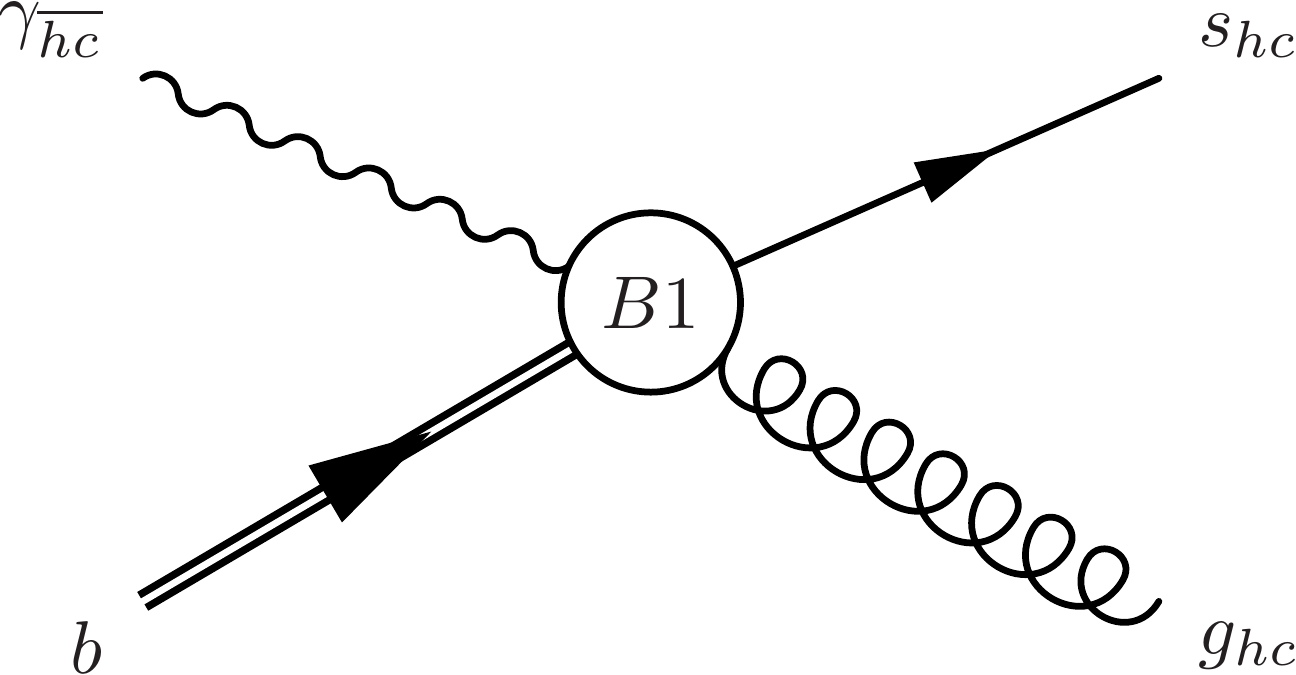}
\caption{\label{fig:SCET} SCET diagrams: direct (left) and  resolved (right) contributions, see text.}
\end{center}
\end{figure}

The first step in the derivation of a factorisation theorem is hard matching. We have to match the electroweak operator onto SCET. The operator under consideration, ${O}_8$~\footnote{For the various conventions we refer  the reader to Ref.~\cite{Hurth:2023paz}}:
\begin{equation}
 {O}_{8g} = -\frac{g_s}{8\pi^2}\,m_b\, \bar s\sigma_{\mu\nu}(1+\gamma_5) G^{\mu\nu} b\,,
\end{equation}  
matches onto two possible SCET operators. 
The direct contribution is represented by a next-to-leading power (NLP) $B$-type current in SCET, i.e. a power-suppressed current composed of two collinear building blocks (see left figure of  Figure~\ref{fig:SCET}):
\begin{equation}
\mathcal{O}_{8g}^{B1}\left(u\right)= \int\frac{dt}{2\pi}e^{-ium_{b}t}\overline{\chi}_{hc}\left(t\bar n\right)\gamma_{\nu\perp}\,  Q_s\,\mathcal{B^{\nu}}_{\overline{hc}\perp}\left(0\right)\gamma_{\mu\perp} \, \mathcal{A^{\mu}}_{hc\perp}\left(0\right)\left(1+\gamma_{5}\right) h\left(0\right),\label{eq:OB1}
\end{equation}
with $Q_s$ as the electric charge of the strange quark in units of $e$ and the electromagnetic gauge-invariant transverse photon field 
\begin{equation}
\mathcal{B^{\nu}}_{\overline{hc}\perp} = e\left(A_{\perp}^{\nu}-\frac{\partial_{\perp}^{\nu}}{n\partial} nA\right).\,
\end{equation}
The second operator in SCET on which the electroweak operator ${O}_{8g}$ matches is a leading-power (LP) $A$-type current, {i.e. a  leading-power current including  one collinear building block, }:
\begin{equation}\label{eq:A0L}
\mathcal{O}_{8g}^{A0}\left(0\right)=\overline{\chi}_{\overline{hc}}\left(0\right)\frac{\slashed n}{2}\gamma_{\mu\perp}\mathcal{A^{\mu}}_{hc\perp}\left(0\right) \left(1+\gamma_{5}\right) h\left(0\right)\,,
\end{equation}
with the gauge invariant gluon field $\mathcal{A^{\mu}}_{hc\perp}=W_{hc}^{\dagger}\left[D_{hc\perp}^{\mu}W_{hc}\right]=\mathcal{A}_{hc\perp}^{a\mu}t^{a}$.
    Then the resolved contribution is represented by a time-ordered product of this  leading-power (LP) $A$-type current,
 with a subleading  $\mathcal{L}_{\xi q}^{\left(1\right)}$ Lagrangian, see right figure of Fig.~\ref{fig:SCET}.
\begin{figure}
\begin{center}
\includegraphics[width=0.4\textwidth]{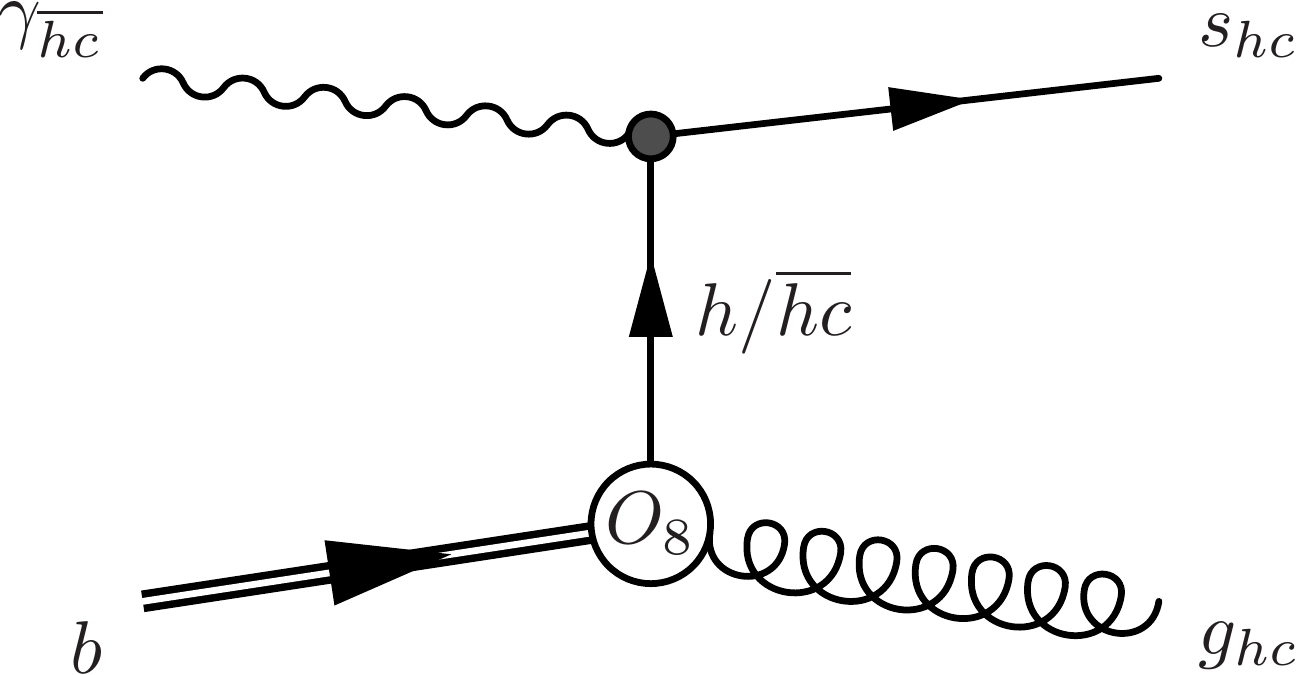}
\caption{\label{fig:QCD} QCD diagram at LO, see text.}
\end{center}
\end{figure}

We can compare the different kinematics of the diagrams with $A$- and $B$-type currents in Figure~\ref{fig:SCET}.  The external s-quark carries hardcollinear momentum. Therefore the intermediate propagator is hard. This situation is represented in SCET by the $B$-type current. When the momentum of the external s-quark tends to zero, the propagator becomes anti-hardcollinear and cannot be integrated out -- it must be reproduced by a dynamical field in the low energy EFT. This situation is represented in SCET by the time-order product of subleading Lagrangian and the $A$-type current. The degeneracy in the EFT description is the reason why the SCET develops divergencies in the convolution integrals. 

For the explicit matching of the two currents at leading order (LO)  we find
\begin{equation}
\label{matchingbLO}
{{\bf \color{ForestBrown} {C_{LO}^{B1}\left(m_{b},u\right)}} = (-1) \frac{\overline{u}}{u}\, \frac{m_{b}^{2}}{4\pi^{2}}\,\frac{G_{F}}{\sqrt{2}}\lambda_{t}\, C_{8g}} = (-1) \frac{\overline{u}}{u} {\bf \color{ForestBrown} C_{LO}^{A0}\left(m_{b}\right)} \,,
\end{equation}
where we use the  hardcollinear momentum fraction $u=\frac{\bar n p_s}{m_{b}}$ and $\overline{u}=1-u = \frac{\bar n p_r}{m_b}$ with the hardcollinear momenta of the strange quark and the gluon, $\bar n p_s$ and $\bar n p_r $  in the direct contribution.

\section{Bare factorisation theorem and endpoint divergences}

The derivation of the factorisation theorem follows the standard approach \cite{Moult:2019mog,Vita:2020ckn, Jaskiewicz:2021cfw}. 
We first perform the soft decoupling transformation~\cite{Bauer:2001yt}, but we do not use a new notation for the hardcollinear fields after decoupling. The decay rate is obtained from the imaginary part of the current-current correlator. The states factorise and thus allow taking matrix elements separately in hardcollinear, anti-hardcollinear and soft sectors. 
{The hardcollinear matrix elements lead to jet functions.  Integration of the anti-hardcollinear fields at the amplitude level leads to antihardcollinear (radiative) jet functions. We can then simplify the Dirac and colour structure, and the remaining soft matrix element defines the shape functions.}\\

{\bf For the direct contribution} we find the bare factorisation theorem on the operator level (see Figure~\ref{fig:FactDirect}
for a schematic description):
\begin{figure}
\begin{center}
\includegraphics[width=0.8\textwidth]{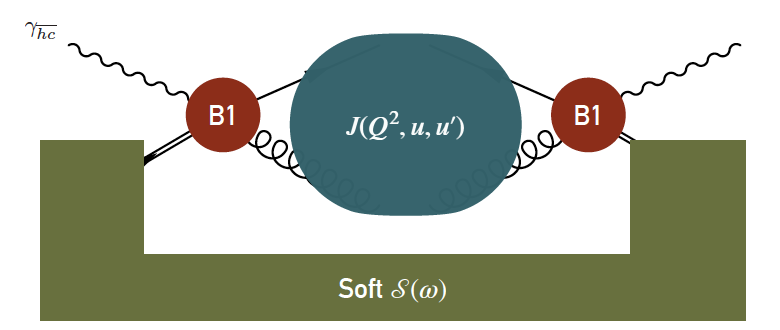}
\caption{\label{fig:FactDirect} Schematic description of the factorisation of the direct contribution, see text.}
\end{center}
\end{figure}
\begin{equation}
\frac{d\Gamma}{dE_{\gamma}}= \mathcal{N}_B\, \int_{0}^{1}du\, {\bf \color{ForestBrown} C^{B1}\left(m_{b},u\right)}\int_{0}^{1}du' {\bf\color{ForestBrown} C^{B1*}\left(m_{b},u'\right) }\int_{-p_{+}}^{\overline{\Lambda}}d\omega {\bf \color{ForestBlue}\,  J\left(M_{B}\left(p_{+}+\omega\right),u,u' \right)} {\bf \color{ForestGreen}  {\cal S}\left(\omega\right)}\label{DIRECT}
\end{equation}
$\mathcal{N}_B$ is a prefactor~\footnote{$\mathcal{N}_B = e^2 Q_s^2 \frac{E_\gamma}{2\pi}$}.
The hard function ${\bf \color{ForestBrown} C^{B1}}$ is given at LO in the last section as a result of the matching of QCD on SCET.
The hardcollinear jet function is a genuine next-to-leading (NLP) object. We define it as a vacuum matrix element of a product of hardcollinear fields:
\begin{align} 
&{\bf \color{ForestBlue} J\left(p^{2},u,u'\right)} = \frac{(-1)}{2 N_c}\, \frac{1}{2\pi} \, \int\frac{dtdt'}{\left(2\pi\right)^{2}}\,\,d^{4}x\,\,e^{-im_{b}\left(ut-u't'\right)+ipx} \label{jetbwithoutopen}\\ 
&\text{Disc}\left[\left\langle 0\right| tr\left[\frac{ 1+\slashed v}{2}(1-\gamma_5) \slashed{\mathcal{A}}_{hc\perp}\left(x\right)  \gamma_{\perp}^{\nu}\chi_{hc}\left(t'\bar{n}+x\right)\overline{\chi}_{hc}\left(t\bar{n}\right)\gamma_{\nu\perp}\slashed{\mathcal{A}}_{hc\perp}\left(0\right) \left(1+\gamma_{5}\right)\right]\left|0\right ] \rangle \right] \,.
\nonumber \end{align}
The soft function - the leading power (LP) shape function - is defined as~\cite{Neubert:1993um}
\begin{align}
{{\bf \color{ForestGreen} {\cal S}\left(\omega\right)}} =\frac{1}{2m_{B}}\int\frac{dt}{2\pi}e^{-i\omega t}\left\langle B\right|h\left(t n\right)S_{n}\left(t n\right)S_{n}^{\dagger}\left(0\right)h\left(0\right)\left|B\right\rangle \,.\label{softleadingpower}
\end{align}
with soft Wilson lines denoted by $S$. 

There is an endpoint divergence in the convolution of the hard matching coefficients and the jet function within the direct contribution. At LO, one finds explicitly with the hard matching coefficient  
${{\bf \color{ForestBrown} {C_{LO}^{B1}\left(m_{b},u\right)}}}$ given in Eq.~\ref{matchingbLO} and the LO jet function given by~\footnote{$A(\epsilon)$ is a function which is finite  for $\epsilon \to 0$.}
\begin{align} 
{\bf \color{ForestBlue}J\left(p^2,u,u'\right) }= C_F \frac{\alpha_s}{4\pi\, m_b}\theta(p^2)\, A(\epsilon)\, \delta(u-u') u^{1-\epsilon} (1-u)^{-\epsilon} \left(\frac{p^2}{\mu^2} \right)^{-\epsilon}\label{JetfunctionMSbar}
\end{align}
that the convolution of both quantities. diverges in the $u \to 0$ limit:
\begin{equation}
\int_0^1 \frac{du}{u}\, \int_0^1 \frac{du'}{u'}  \delta(u-u') u^{1-\epsilon} \approx \int_0^1 \frac{du}{u^{1+\epsilon}}\,.
\end{equation} 
Consequently, due to endpoint divergences, the bare factorisation formula of the direct contribution is already invalid for the $d\to 4$ limit at LO. \\

\begin{figure}
\begin{center}
\includegraphics[width=0.8\textwidth]{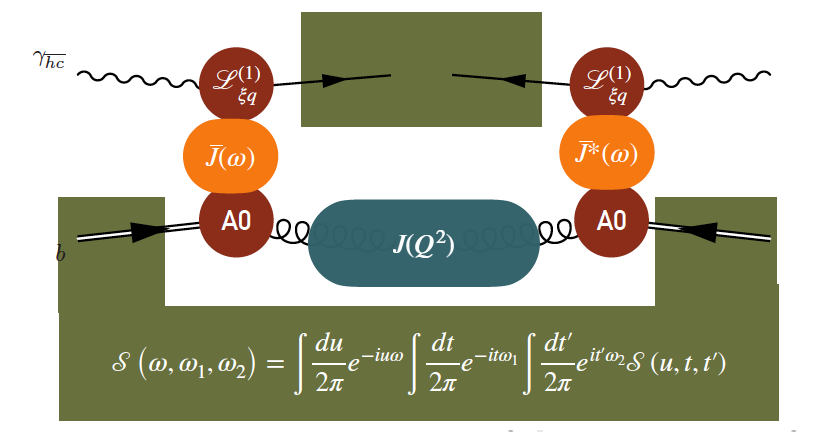}
\caption{\label{fig:FactResolved} Schematic description of the factorisation of the resolved contribution, see text.}
\end{center}
\end{figure}

{\bf For the resolved contribution} we find the following factorisation theorem on the operator level (for a schematic description of the factorisation of the resolved contribution see Figure~\ref{fig:FactResolved}):
\begin{equation}\label{RESOLVED}
\frac{d\Gamma}{dE_{\gamma}} = \mathcal{N}_A \left| {\color{ForestBrown}\bf C^{A0}\left(m_{b}\right)} \right|^{2} { \int_{-p_+}^{\overline{\Lambda}}} \hspace{-0.4cm} d\omega\,  {{\color{ForestBlue}\bf J_{g}\left({ m_{b}}\left(p_{+}+\omega\right)\right)}}
\int\hspace{-0.2cm} d\omega_{1}\hspace{-0.2cm}\int d\omega_{2}\, {\color{ForestOrange} \bf \overline{J}\left(\omega_{1}\right)\, \overline{J}^{*}\left(\omega_{2}\right)}\, {\cal S}\left(\omega,\omega_{1},\omega_{2}\right),
\end{equation}
with the prefactor  $\mathcal{N}_A = \mathcal{N}_B \equiv \mathcal{N}$. The hard function 
${\bf \color{ForestBrown} C^{A}}$ is given at LO in the last section as a result of the matching of QCD on SCET again.
In the hardcollinear sector, there are only gluon fields, so the standard LO gluon jet function ${\color{ForestBlue}\bf J_g(p^2)}$  appears. The anti-hardcollinear jet function is defined on the amplitude level because there is no energetic particle emitted in the anti-hardcollinear directions besides the photons:
\begin{align} 
\mathcal{O}_{T\xi q} &= i\int d^{d}xT\left[ \mathcal{L}_{\xi q}\left(x\right),\mathcal{O}_{8g}^{A0}\left(0\right)\right] 
\nonumber\\
    &= \int d\omega\int\frac{dt}{2\pi}e^{-it\omega}\left[\overline{q_{s}}\right]_{\alpha}\left(t n\right)
    {\color{ForestOrange} \bf \left[\overline{J}\left(\omega\right)\right]_{\alpha\beta}^{a\, \nu\mu} } Q_s\, 
\mathcal{B^{\nu}}_{\overline{hc}\perp}\left(0\right)\mathcal{A}_{hc\perp}^{\mu\, a}\left(0\right)\left[h\left(0\right)\right]_{\beta}.
\end{align}
One can  prove that the antihardcollinear jet function has a unique structure to all orders in $\alpha_s$:
\begin{equation}\label{eq:Jbar}
{{\color{ForestOrange}\bf  \left[\overline{J}\left(\omega\right)\right]_{\alpha\beta}^{a\, \nu\mu} }}={{\color{brown} \overline{J}\left(\omega\right)\,}} t^a\, \left[\gamma_{\perp}^{\nu}\gamma_{\perp}^{\mu}\frac{\slashed {\bar n}\slashed n}{4}\right]_{\alpha\beta},
\end{equation}
Finally, the soft function is given by 
\begin{align} \label{SOFT}
{\color{ForestGreen}\bf {\cal S}\left(u,t,t'\right)} & ={(d-2)^2}g_{s}^{2}\left\langle B\right|\overline{h}\left(un\right)\left(1-\gamma_{5}\right)  \left[S_{n}\left(un\right)t^{a}S_{n}^{\dagger}\left(un\right)\right]S_{\bar n}\left(un\right)S_{\bar n}^{\dagger}\left(t' \bar n+u n\right) \\ &\frac{\slashed n \slashed {\bar n}}{4}     q_{s}\left(t' \bar n +un\right) 
 \overline{q}_{s}\left(t \bar n\right)\frac{\slashed {\bar n}\slashed n}{4}S_{\bar n}^{}\left(t \bar n \right)S_{\bar n}^{\dagger}\left(0\right)\left[S_{n}\left(0\right)t^{a}S_{n}^{\dagger}\left(0\right)\right]\left(1+\gamma_{5}\right)h\left(0\right)\left|B\right\rangle\, / \, (2 m_B)\nonumber \,. \label{SOFT}
\end{align}
The Fourier transform is defined as 
\begin{equation}
{\color{ForestGreen} \bf {\cal S}\left(\omega,\omega_{1},\omega_{2}\right)}=\int\frac{du}{2\pi}e^{-iu\omega}\int\frac{dt}{2\pi}e^{-it\omega_{1}}\int\frac{dt'}{2\pi}e^{it'\omega_{2}}{\color{ForestGreen}\bf {\cal S}\left(u,t,t'\right)} \,.
\end{equation}

At LO the factorization formula is well behaved as long as $\omega_1,\omega_2  \sim  \omega$ : 
\begin{equation}
\frac{d\Gamma}{dE_{\gamma}} = 
2\mathcal{N}_A\, \left|{{\bf \color{ForestBrown}C_{LO}^{A0}\left(m_{b}\right)}}\right|^{2}\hspace{-0.2cm}
\int^{\overline{\Lambda}}_{-p_+} \hspace{-0.4cm}
d\omega\, \,{{\bf \color{ForestBlue} \delta\left({m_{b}}\left(p_{+}+\omega\right)\right)}} \int_{-\infty}^{\infty} \hspace{-0.4cm} d\omega_{1}\int_{-\infty}^{\omega_1} \hspace{-0.5cm} d\omega_{2}\, {{\color{ForestOrange} \frac{1}{(\omega_1 - i\epsilon)}  \frac{1}{(\omega_2 + i\epsilon)}}} {{\color{ForestGreen}\bf  {\cal S}\left(\omega,\omega_{1},\omega_{2}\right) }}.\label{factorisationA}
\end{equation}
However, for $\omega_{1,2} \gg \omega$,  the soft function can be shown to be asymptotically constant, which leads to endpoint divergence in the convolution integrals of jet and shape functions for large $\omega_{1,2}$. 

In this limit, the soft strange quarks become hardcollinear and can be decoupled from the soft gluons. This way, the soft function reduces to the LP shape function, which also occurs in the direct contribution. This can be shown explicitly taking into account that the $\omega_{1,2} \to \infty$ limit corresponds to the $t,t' \to 0$ limit in Eq.~\ref{SOFT}. In fact, the soft function in the limit  $\omega_{1,2}$, denoted by ${\color{ForestGreen} \bf \widetilde{{\cal S}}}$, can be matched on the LP shape function
with a perturbative kernel $K$:
\begin{align}\label{eq:matchingS}
{\color{ForestGreen} \bf \widetilde{{\cal S}}\left(\omega,\omega_{1},\omega_{2}\right)} = \int d\omega' K(\omega,\omega',\omega_1,\omega_2) {\color{ForestGreen}\bf  {\mathcal S}(\omega')}\,.
\end{align}
At LO, we find:
\begin{equation}
{\color{ForestGreen}\bf \widetilde{{\cal S}}\left(\omega,\omega_{1},\omega_{2}\right)} = C_F A(\epsilon) \frac{\alpha_s}{(4\pi)} \,\, \omega_1^{1-\epsilon} \delta(\omega_1-\omega_2)
\int_\omega^{{\overline\Lambda}} d \omega' \,{\color{ForestGreen}\bf {\cal S}(\omega')} \, \left(\frac{(\omega'-\omega)}{\mu^2} \right)^{-\epsilon}\,,
\label{asymptoticsoftfunction}
\end{equation}
which includes the leading power shape function {\color{ForestGreen} \bf ${\cal S}(\omega)$}.

This makes the endpoint divergence in the convolution integral of jet and shape function within the resolved contribution in the asymptotic limit $\omega_{1,2} \to \infty$
manifest. In this limit, the resolved contribution at LO can be written as: 
\begin{align}
\frac{d\Gamma}{dE_{\gamma}}|^{\rm asy}_A &=
2\mathcal{N}\,{\color{ForestBrown}\bf |C_{LO}^{A0}(m_{b})|^{2}}\hspace{-0.2cm}     \int_{-p_+}^{\overline{\Lambda}}
\hspace{-0.4cm} d\omega {\color{ForestBlue}\bf J_g^{LO}(m_{b}(p_{+}+\omega))}\hspace{-0.2cm} \int_{m_b}^{\infty}  \hspace{-0.4cm} d\omega_{1} {\color{ForestOrange} \bf \overline{J}_{LO}(\omega_1)}\int_0^{\omega_1} \hspace{-0.5cm} d\omega_{2}{\color{ForestOrange}\bf \overline{J}^*_{LO}(\omega_2)}\,\,{\color{ForestGreen}\bf \widetilde{{\cal S}}(\omega,\omega_{1},\omega_{2})}\nonumber\\
&=\mathcal{N} {\color{ForestBrown}\bf |C_{LO}^{A0}\left(m_{b}\right)|^{2}} \,
 \frac{\alpha_sC_F}{(4\pi)\, m_b} \,  \frac{1}{\epsilon} A(\epsilon)\,  \int_{-p_+}^{\overline{\Lambda}}  \hspace{-0.3cm}    d \omega \,{\color{ForestGreen}\bf {\cal S}_{LO}(\omega')}
 \left(\frac{m_b (\omega+p_+)}{\mu^2} \right)^{-\epsilon}\,.
  \label{Acurrentasymptotic}
\end{align}

\section{Refactorisation to all orders in $\alpha_s$}

Evaluating the direct contribution in the asymptotic limit $u,u' \to 0$ (before integration)
\begin{align}
\frac{d\Gamma}{dE_{\gamma}} |^{u,u'\to 0}_B 
&=-\mathcal{N}\, {\color{ForestBrown}\bf \left|C_{LO}^{A0}\left(m_{b}\right)\right|^{2}} \,
 \frac{\alpha_sC_F}{(4\pi)\, m_b} \, \frac{1}{\epsilon}\, A(\epsilon) \int_{-p_+}^{\overline{\Lambda}}     d \omega\, 
 {\color{ForestGreen}\bf {\cal S}_{LO}(\omega) }
\left(\frac{m_b (\omega+p_+)}{\mu^2} \right)^{-\epsilon}\,, \label{Bcurrentasymptotic}
\end{align} 
and comparing with the asymptotic limit ($\omega_1,\omega_2 \to \infty$) of the resolved contribution given in 
Eq.~\ref{Acurrentasymptotic} one verifies at LO 
\begin{equation} \label{CruxLO}
\frac{d\Gamma}{dE_{\gamma}}|^{\rm asy}_A = (-1) \frac{d\Gamma}{dE_{\gamma}} |^{u,u'\to 0}_B  \,.
\end{equation} 
Thus, the sum of the two terms is finite and equal to zero, and the endpoint divergences cancel. This reflects the fact that in the limits $u \to 0$ and $\omega_{1}\sim\omega_{2}\gg\omega$ the two terms of the subleading ${\cal O}_8 - {\cal O}_8$ contribution have the same structure. This LO result is a special case of the all-order relation which can be formulated at the operator level.\\ 

One can derive three refactorisation conditions which reflect this fact  that 
in the limits \mbox{$u\sim u' \ll 1$} and 
$\omega_{1}\sim\omega_{2}\gg\omega$ the resolved and the direct contribution have the same structure to all orders in $\alpha_s$. The refactorisation relations are operatorial relations that then guarantee the cancellation of endpoint divergences between the two terms to all orders in $\alpha_s$ as we will see below. 
\begin{itemize}
\item We find that in the limit $u\to0$, the matching coefficient can be further factorised 
\begin{equation}
{{\left\llbracket {\color{ForestBrown}\bf  C^{B1}\left(m_{b},u\right)} \right\rrbracket =(-1) {\color{ForestBrown}\bf C^{A0}\left(m_{b}\right)} \,m_b \,{\color{ForestOrange} \bf \overline{J}\left(u m_b \right)\,, \label{eq:RF1}}}}
\end{equation}
where $\left\llbracket g(u) \right\rrbracket$ only denotes the leading term of a function $g(u)$ in the limit $u \to 0$ 
and without any higher power corrections in $u\ll 1$. 
\item We find the new soft function 
{\color{ForestGreen}\bf  $\widetilde{{\cal S}} \left(\omega,\omega_{1},\omega_{2}\right)$} which corresponds to the function {\color{ForestGreen} \bf ${\cal S}\left(\omega,\omega_{1},\omega_{2}\right)$}
in the limit $\omega_{1}\sim\omega_{2}\gg\omega$.   
In this limit, we can consider the light soft quarks to be hardcollinear. In this function 
{\color{ForestGreen} \bf $\widetilde{{\cal S}}\left(\omega,\omega_{1},\omega_{2}\right)$} higher power corrections in         $\omega/\omega_{1,2}$ are neglected. 
\item We find that the jet function  {\color{ForestBlue} \bf $J \left(m_{b}\left(p_{+}+\omega\right),u,u'\right)$} fulfills the following relation
in the limit $u \to 0$ and $u' \to 0$:
\begin{equation}  \label{refactsquared}
\int_{-p_+}^{\overline{\Lambda}} \hspace{-0.4cm} d\omega
\left\llbracket {\color{ForestBlue}\bf J \left(m_{b}\left(p_{+}+\omega\right),u,u'\right)} {\color{ForestGreen} \bf {\cal S}(\omega)}\right\rrbracket =
 \int_{-p_+}^{\overline{\Lambda}} \hspace{-0.4cm}d\omega
{\color{ForestBlue}\bf J_{g}(m_b(p_{+}+\omega))}
{\color{ForestGreen} \bf \widetilde{\cal S}(\omega,m_b u ,m_b u')}\,,
\end{equation}
where the brackets indicate that the $u \to 0$ {\it and} $u' \to 0$ limits have to be taken. \\
\end{itemize}
\vspace{-0.5cm}
Using these all-orders refactorisation conditions we then can  cast the two asymptotic subtraction terms into the following form:
 \begin{align}
0=\:&2\mathcal{N}\left|{\color{ForestBrown}\bf C^{A0}}\left(m_{b}\right)\right|^{2}\int_{-p_{+}}^{\overline{\Lambda}}d\omega {\color{ForestBlue}\bf J_{g}\left(m_{b}\left(p_{+}+\omega\right)\right)}\int_{m_{b}}^{\infty}d\omega_{1}{\color{ForestOrange}\bf \overline{J}\left(\omega_{1}\right)}\, \int_{0}^{\omega_1}d\omega_{2}{\color{ForestOrange}\bf \overline{J}^*\left(\omega_{2}\right)}{\color{ForestGreen}\bf \widetilde{{\cal S}}\left(\omega,\omega_{1},\omega_{2}\right)}   \nn\\
 +\:& 2\mathcal{N}\int_{0}^{1} \hspace{-0.3cm} d u   \left\llbracket {\color{ForestBrown}\bf C^{B1}\left(m_{b},u\right)}\right\rrbracket    \int_{u}^{1}\hspace{-0.3cm} d u'  \left\llbracket {\color{ForestBrown}\bf C^{B1*}\left(m_{b},u'\right)}\right\rrbracket  \int_{-p_{+}}^{\overline{\Lambda}}\hspace{-0.3cm}d\omega    \left\llbracket {\color{ForestBlue} \bf J\left(m_{b}\left(p_{+}+\omega\right),u,u'\right)}{\color{ForestGreen}\bf  {\cal S}(\omega)}\right\rrbracket \:.  \label{Bsubtraction3}
 \end{align}
We then subtract these asymptotic terms from the sum of the two all-order bare factorisation theorems for resolved and direct contribution  derived in the last section (see Eqs.~\ref{DIRECT} and \ref{RESOLVED}) and obtain the endpoint finite factorisation theorem. 
\begin{align}\label{eq:bareFT}
\frac{d\Gamma}{dE_{\gamma}}|_{A+B} &= 2 \mathcal{N} \int_{-p_{+}}^{\overline{\Lambda}} 
 \hspace{-0.1cm}   d\omega \bigg\{  {\color{ForestBlue}\bf  J_g(m_{b}(p_{+}+\omega))} \left|{\color{ForestBrown} \bf C^{A0}\left(m_{b}\right)}\right|^{2}   \\
&\times \int_{-\infty}^{\infty}  \hspace{-0.1cm}   d\omega_{1} \,  \int_{-\infty}^{\omega_1}  \hspace{-0.1cm}  d\omega_{2} {\color{ForestOrange}\bf \overline{J}(\omega_1)\,\overline{J}^*(\omega_2) }\left[{\color{ForestGreen}\bf {\cal S}\left(\omega,\omega_{1},\omega_{2}\right)}\nonumber- \theta(\omega_1-m_b)\theta(\omega_2)
{\color{ForestGreen} \bf \widetilde{\cal S}(\omega,\omega_{1},\omega_{2})}\right]\nonumber\\
&+ \int_{0}^{1} \hspace{-0.1cm}   du\int_{u}^{1} \hspace{-0.1cm}  du'  \, \Big[ {\color{ForestBrown}\bf C^{B1}\left(m_{b},u\right)\,
C^{B1*}\left(m_{b},u'\right)}\,{\color{ForestBlue}\bf J\left(m_{b}\left(p_{+}+\omega\right),u,u'\right)} {\color{ForestGreen}\bf {{\cal S}}\left(\omega\right)}\nonumber\\
&-   \left\llbracket {\color{ForestBrown}\bf C^{B1}\left(m_{b},u\right)}
\right\rrbracket  
 \left\llbracket {\color{ForestBrown}\bf C^{B1*}\left(m_{b},u'\right)}\right\rrbracket\left\llbracket {\color{ForestBlue}\bf J\left(m_{b}\left(p_{+}+\omega\right),u,u'\right)}{\color{ForestGreen}\bf {\cal S}(\omega)}\right\rrbracket  
 \Big]\bigg\}\:,\nonumber
\end{align}
At this point, the convolutions integrals in the $A$- and $B$-type contributions are no longer divergent, and we can renormalise the functions entering the factorisation theorem and take the limit $d\to4$.  In a final step, we are able to show  that endpoint reshuffling and renormalisation effectively commute~\cite{Hurth:2023paz}.

\vspace{2cm}

{\bf Acknowledgements }  TH is  supported by  the  Cluster  of  Excellence  ``Precision  Physics,  Fundamental
Interactions, and Structure of Matter" (PRISMA$^+$ EXC 2118/1) funded by the German Research Foundation (DFG) within the German Excellence Strategy (Project ID 390831469), 
RS is supported by the United States Department of Energy under Grant Contract DESC0012704. R.S. and TH thank the organisers of SCET23 and FPCP23 for their excellent organisation and generous hospitality.

\end{document}